# Does Global seismic energy release increase? An analysis based on the Lithospheric Seismic Energy Flow Model (LSEFM). The case of mega - earthquakes (M > 9)

Thanassoulas[1], C., Klentos[2], V., Verveniotis[3], G., Zymaris, N.[4]

1. Retired from the Institute for Geology and Mineral Exploration (IGME), Geophysical Department, Athens, Greece.
   e-mail: thandin@otenet.gr - URL: www.earthquakeprediction.gr

2. Athens Water Supply & Sewerage Company (EYDAP),
   e-mail: klenvas@mycosmos.gr - URL: www.earthquakeprediction.gr

3. Retired Ass. Director, Physics Teacher at 2nd Senior High School of Pyrgos, Greece.
   e-mail: gver36@otenet.gr - URL: www.earthquakeprediction.gr

4. Retired, Electronic Engineer.

**Abstract**

In this work the data of the earthquake catalog of the NOAA, National Geophysical Data Center (NGDC) are processed in terms of global seismic energy release. The determined Global Cumulative Seismic Energy Release (GCSER) graph as a function of time, is analyzed in the magnitude domain (discrete energy windows). Characteristic components of the analyzed graph are: its accelerated deformation character observed for energy windows lower than the background seismicity (M = 7.0 – 7.5), its "lock" state that started on 1923 and long seismic quiescence periods that preceded recent mega - earthquakes. The background GCSER value oscillates during the last century with a period of 60 years and with increasing amplitude. The recent (1952 – 2012) 5 mega - earthquakes are closely related to the amplitude increase of the GCSER oscillation. Hence, it is suggested that more mega - earthquakes are probable due to occur in the future. A global mechanism is postulated for the generation of the mega - earthquakes based on the principles of the non-linearity of the earth, the Markovitz wobble and the expanding earth. The determination of the magnitude of the 5 recent mega - earthquakes by the Lithospheric Seismic Energy Flow Model (LSEFM) method indicates that the latter is globally applicable.

## 1. Introduction.

The generation of an earthquake is the result of localized stress – strain accumulation in the focal area. The occurrence of an earthquake changes the stress – strain field of the regional seismogenic area but physical mechanisms for stress and strain redistribution in the lithosphere for long distances are not known yet. Large and global scale mechanisms that take place in the lithosphere have been proposed by researchers (Barenblatt et al., 1983; Press and Allen, 1995). Another approach regarding long range earthquake correlation is based on the common feature of complex non-linear hierarchical dynamic systems (Keilis-Borok, 1990, 2002; Sornette and Sammis, 1995; Turcotte et al., 2000), of which the lithosphere is a typical example.

Regarding the preparation area of an earthquake, it has been found that it exceeds at least by ten times the rupture length (Dobrovolsky et al., 1979; Sadovsky, 1986; Keilis-Borok, 1990, 2002; Press and Allen, 1995; Bowman et al. 1998). Following that rupture length – preparation area size relation the mega – earthquake of Sumatra (December 26, 2004, M = 9.0) theoretically might have involved an area of about 15000 Km in linear dimension. Since this dimension exceeds the diameter of the earth, the generation of such a large earthquake suggests a global interaction and not a local even large faulting system (Romashkova and Kossobokov, 2007).

Consequently, it is interesting to search for premonitory phenomena, by analysing the lithosphere as a single whole, which represents the ultimate scale of the complex Earth's hierarchy. Global seismicty studies have been reported by many researchers (Benioff, 1951; Mogi, 1979; Romanowicz 1993; Bufe and Perkins, 2005) who studied the seismic cycle and identified some global scale patterns in seismic energy release over decades. The general global approach to the Earth seismic dynamics naturally arises from the conception of the Earth lithosphere as a complex non-linear dynamic system consisting of hierarchy of interacting different-scale blocks (Keilis-Borok, 1990, 2002; Keilis-Borok and Soloviev, 2003). The lithosphere as a whole represents an ultimate member of this hierarchy.

Romashkova and Kossobokov (2007) suggested in the frame of a global seismicity study, that are quite evident "(i) *the presence of global scale intermediateterm tectonic processes in the lithosphere*; (ii) *the occurrence of the global scale seismicity patterns implying criticality in the last decade. The December 26, 2004 Sumatra-Andaman mega-earthquake may represent either the culmination or one of the successive stages of such a process in the last decade*".

In this work, for studying the global seismicity, we will apply the Lithospheric Seismic Energy Flow Model (LSEFM), introduced by Thanassoulas (2007), and the entire lithosphere will be considered as the "open physical system" that absorbs strain energy and at the same time it releases seismic energy too. It will be investigated whether mega - earthquakes (M > 9) of the last 100 years comply with the physical mechanism of LSEFM and whether their magnitude could be determined well in advance of their occurrence by analyzing the seismic energy release over the last century.

## 2. The physical working model.

The regional area of the lithosphere where the EQ will occur is considered as an open physical system that absorbs and releases seismic energy. Practically, the lithosphere is at a state of critical dynamic equilibrium, or it is teetering on the edge of instability, with no critical length scale (Bak, Tang and Wiesenfeld 1988; Bak, 1996) as far as it concerns in-flow and out-flow of strain energy. The following figure (1) shows the postulated "lithospheric seismic energy - flow model" in such a condition.

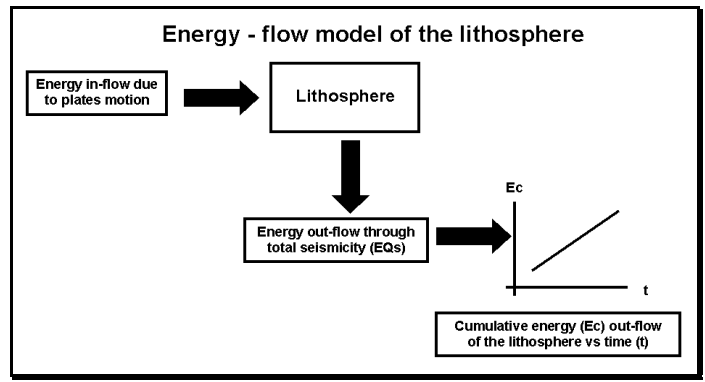

Fig. 1. Energy – flow model of the lithosphere and cumulative seismic energy out-flow (Ec), as a function of time (after Thanassoulas, 2007, 2008).

Theoretically, following the above working physical model, the cumulative energy release in a seismogenic area, during a normal, seismic period, should be a linear function (Thanassoulas, 2008) of time, due to energy conservation law of Physics. This is demonstrated in the lower right part of the figure (1). Deviations from the normal state lead to:

a. Accelerating deformation state.

b. "Lock" seismic state due to decrease of seismic energy release (seismic quiescence) and hence continuous charging of the focal area with strain load.

Examples for (a) and (b) are shown below in figures (2 and 2a).

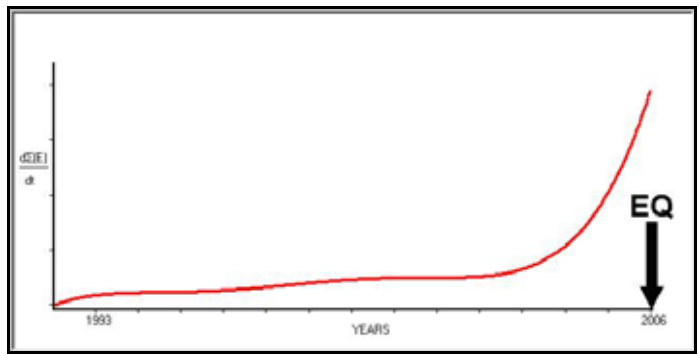

Fig. 2. Increased, accelerated, seismic energy release rate (or accelerated deformation), observed prior to the EQ, East of Kythira, Greece, 08/01/2006, Ms = 6.9R.

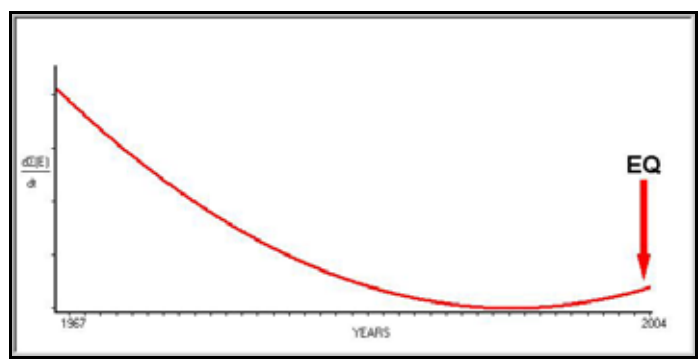

Fig. 2a. De-accelerating, seismic energy release rate (seismic quiescence), observed, for the EQ of Lefkada, Greece, 14/8/2003, Ms = 6.4R.

The above presented model will be applied for testing purposes on data related to global seismicity.

## 3. The data.

The earthquake data that were used in this work is the catalog provided by the NOAA National Geophysical Data Center (NGDC). The first part of the catalog refers to historical earthquakes (-2150 to 1600) and obviously is not complete for any use. From 1600 to 1850, although the data are more "dense", still are not complete too. The situation changes drastically when the instrumental seismology was introduced about 150 years ago.

Consequently, in the present study we will use only the part of the catalog that is objectively complete (1900 – 2012).

## 4. Considering a global lithospheric physical model.

As an example of the change of the recorded global seismicity along time of the NGDC catalog the following diagram of the corresponding global cumulative seismic energy release is presented in figure (3) that corresponds to the period 1000 – 2012.

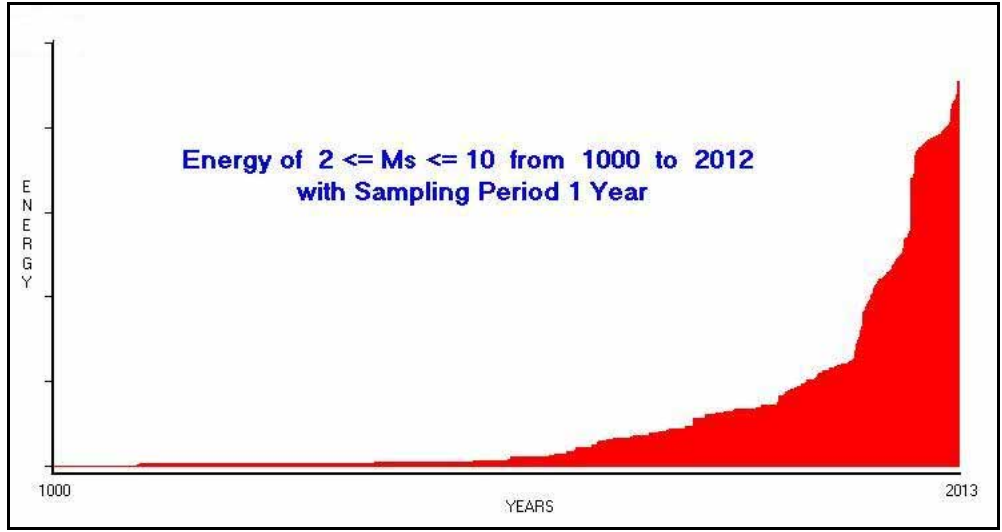

**Fig. 3.** Change of the recorded global cumulative seismic energy release along time of the NGDC catalog that corresponds to the period 1000 – 2012. The sampling interval is 1 year.

An inspection of figure (3) reveals that during the time period from 1000 to 1500 no significant global seismic energy release is observed due to the catalog incompleteness. From 1500 to 1800 a slight increase is observed while the most drastic change start happening on 1900 due to the introduction of the instrumental seismology. Thus, the present analysis will be based on the most reliable part of the earthquake catalog that covers the 1900 – 2012 period of time. The latter, in terms of global cumulative seismic energy release, is presented in the following figure (4). A sampling interval of 1 month has been used in this case.

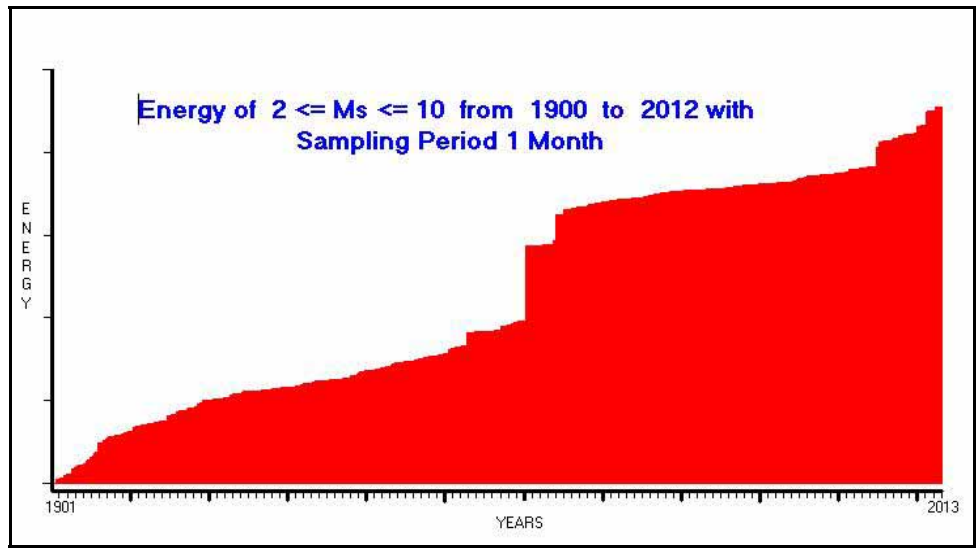

**Fig. 4.** Global cumulative seismic energy release (GCSER) is shown that corresponds to the period of 1900 to 2012. The sampling interval is 1 month.

The form of figure (4) shows that there are periods of time with increased seismicity, comparatively decreased seismicity and sudden steps of it that correspond to large seismic events. Since the global cumulative seismic energy release is a function of time, it could be analyzed by typical spectral analysis packages in to its basic frequency components.

$$GCSER(t) = F(f_1, f_2, \ldots f_n, t) \quad (1)$$

where $f_1, f_2, \ldots f_n$ are the different frequency components of GCSER(t).

In terms of seismology, the global cumulative seismic energy release is a time function that strictly depends on the magnitudes of the occurred earthquakes. Consequently, equation (1) can take the form of:

$$GCSER(t) = f(M_1, M_2, \ldots M_n, t) \quad (2)$$

where $M_1, M_2, \ldots M_n$ are the different magnitude components of GCSER(t).

Therefore, it is interesting to analyze figure (4) in terms of earthquake magnitudes.

The magnitude window for the present analysis was increased in steps of dM = 0.5 from M = 4.0 to M = 9.0
At first, a rather small magnitude has been selected of M = 4.0 – 4.5 that is the magnitude lower threshold of the earthquake catalog. The corresponding global cumulative seismic energy release is shown in figure (5).

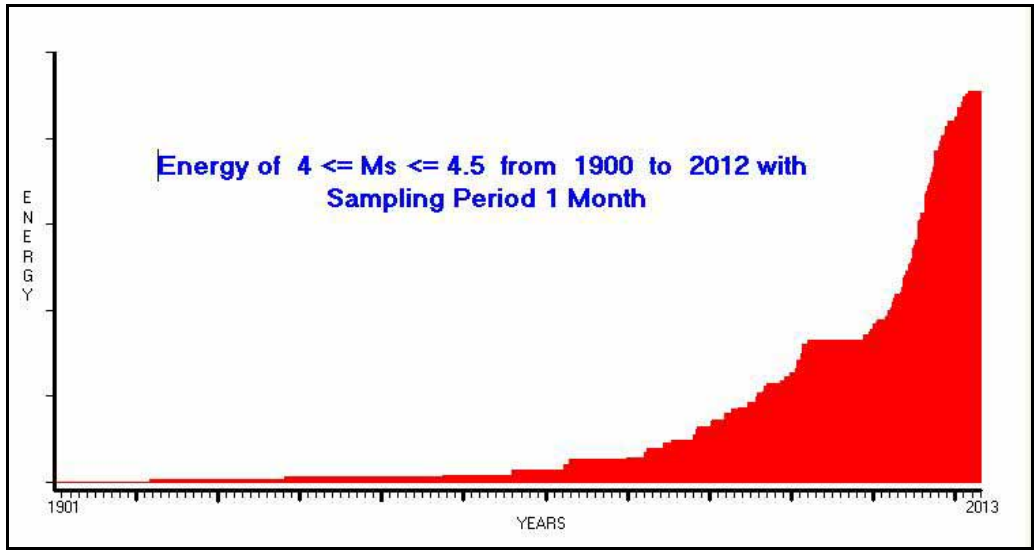

**Fig. 5. Global cumulative seismic energy release for M = 4.0 – 4.5.**

The form of that graph is the typical one of the accelerated deformation. The latter could be attributed either to a real increase of the observed seismicity, that in turn could suggest an oncoming large seismic event, or by the fact that the seismological observatories have been improved technically and more seismic small events have been registered in the recent decades of years.

For magnitude windows up to M = 7.0 the form of the global cumulative seismic energy release gradually approaches the one of next figure (6) that corresponds to a magnitude window of M = 7.0 – 7.5.

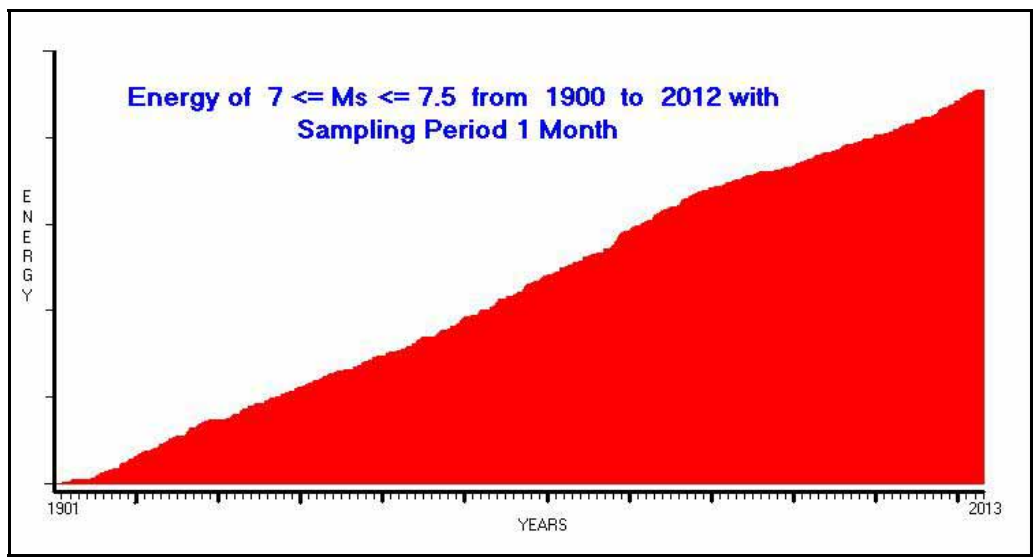

**Fig. 6. Global cumulative seismic energy release for M = 7.0 – 7.5.**

The global cumulative seismic energy release for the magnitude of M = 7.0 – 7.5 is resembled closely by a straight line, therefore, the specific magnitude window indicates indirectly what seismicity magnitude represents the normal background global lithospheric conditions. A similar analysis (Thanassoulas et al. 2010) had indicated that a magnitude of M = 4.5 is valid for the normal background seismicity for the Greek territory.

At the next magnitude window shown in figure (7) for M = 7.5 – 8.0, the global cumulative seismic energy release conditions have slightly changed. At the start of the century an increased rate of cumulative seismic energy release is observed that is followed by a stable and lower one for the rest of the period o time.

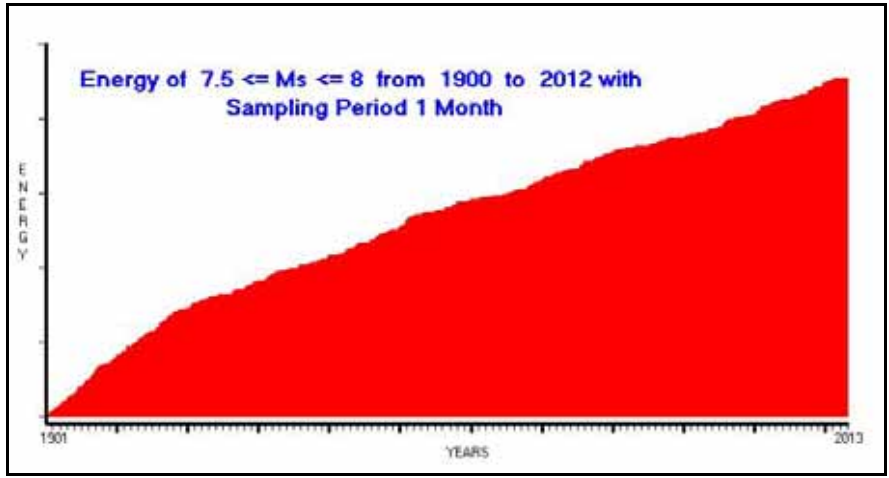

**Fig. 7. Global cumulative seismic energy release for M = 7.5 – 8.0**

The latter could possibly be interpreted as a lengthy seismic quiescence period that leads towards future large seismic events. The next magnitude window (M = 8.0 – 8.5) shown in figure (8) looks more rough due to larger used earthquake magnitudes.

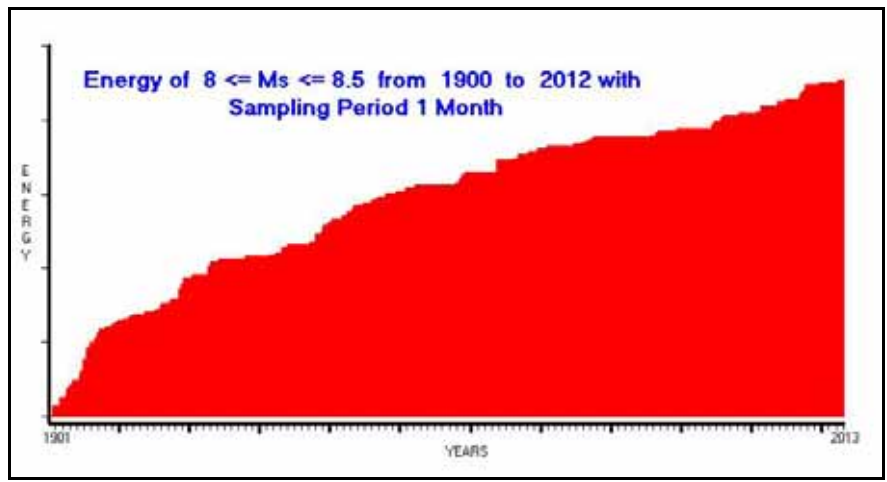

**Fig. 8. Global cumulative seismic energy release for M = 8.0 – 8.5**

The effect of the larger magnitude and infrequent earthquakes is more obvious in figure (8).
Finally, the magnitude window of M = 8.5 – 9.0 is shown in figure (9).

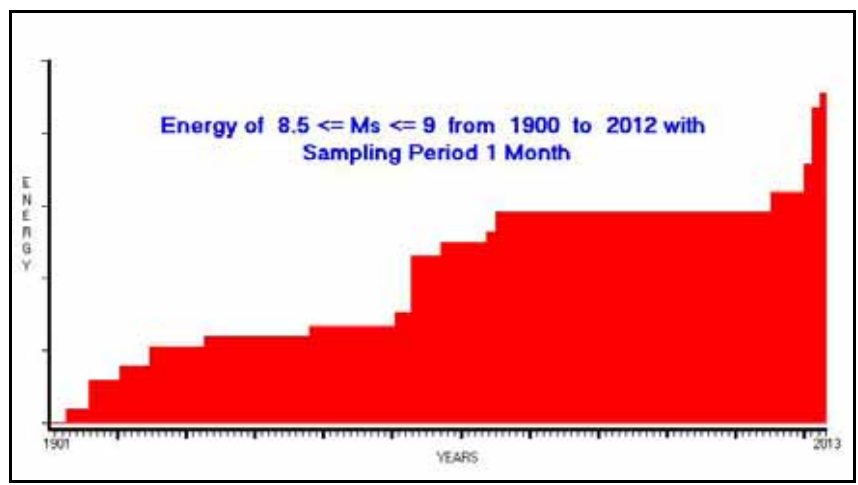

**Fig. 9. Global cumulative seismic energy release for M = 8.5 – 9.0**

In the case of figure (9) characteristic periods of time exist which are distinguished by no increase of the global cumulative seismic energy release. The latter is obvious since large earthquakes do not occur quite often in time. Periods of intense and large seismicity are shown by large steps of the released global cumulative seismic energy.

Let us recall figure (6) that represents the global background seismic energy release and hence the corresponding background seismicity. The fit of the GCSER to a straight line is presented in the following figure (10).

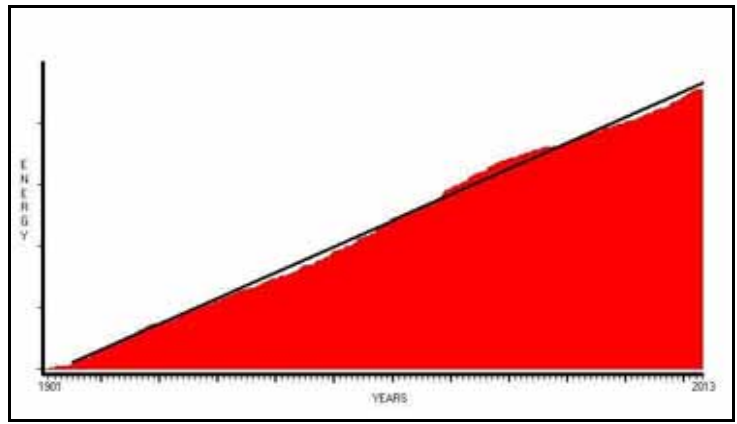

**Fig. 10. Actual GCSER, for M = 7.0 – 7.5, fitted by a straight line that represents its background level.**

It is evident from figure (10) that the GCSER shows an oscillating character for the last 70 years. In order to get a clearer picture of this phenomenon, a $7^{th}$ order polynomial has been fitted to the data of figure (10). The actual GCSER data are compared to the ones generated by the fitted polynomial in the following figure (11).

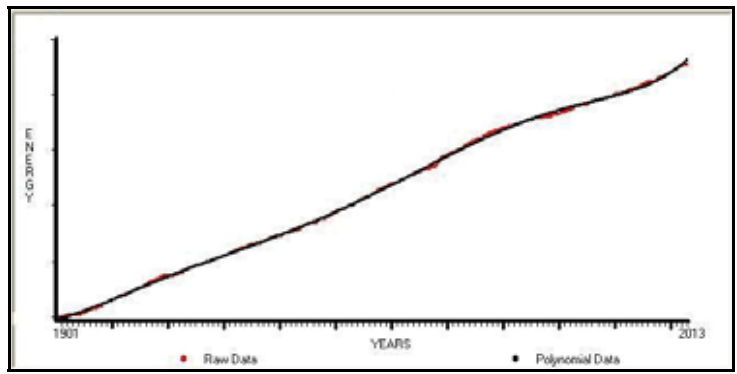

**Fig. 11. GCSER data (red line) compared to the ones generated by a $7^{th}$ order fitted polynomial function (black line).**

Consequently, after having determined the form and the parameters of the fitted polynomial function, the rate of change in time of the GCSER is calculated by analytical derivation, and presented in the following figure (12).

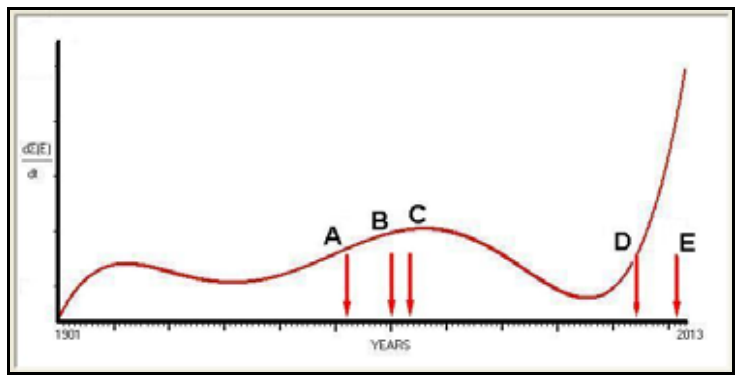

**Fig. 12. Rate of change in time of the GCSER is shown. Mega - earthquakes (A, B, C, D, E with M>9.0) that occurred in this period of time are indicated by vertical red arrows.**

During 1900 to 2012 five (5) mega - earthquakes occurred in a global scale. The following table (1) indicates their: occurrence time (year), geographical coordinates and magnitude. Actually, these five (5) mega - earthquakes are the only ones registered in the NGDC catalog from -2150 to 2013 and all occurred in the last 100 years.

## TABLE – 1

| EQ | Year | Lat | Lon | M |
|---|---|---|---|---|
| A | 1952 | 52.75 | 159.50 | 9.0 |
| B | 1960 | -39.50 | -74.50 | 9.5 |
| C | 1964 | 61.04 | -147.73 | 9.2 |
| D | 2004 | 3.30 | 95.98 | 9.1 |
| E | 2011 | 38.30 | 142.37 | 9.0 |

An interesting feature of figure (12) is its oscillating character. The period of the oscillation ranges between 50 – 60 years depending on what kind (maxima or minima) of successive peaks are selected for the period determination. Another interesting feature is its continuous increase of oscillation amplitude. Increased rate of change of oscillating GCSER values indicates increased oscillating cumulative seismic energy release. The latter indicates in turn increased oscillating seismicity, which in turn means increased oscillating stress – strain global accumulation. The earthquakes designated with A, B, C occurred during a specific increase of the GCSER (1952 – 1964) while the earthquakes designated with D, E occurred during the last decade of years (2004 – 2011) and specifically during the last increasing branch of the GCSER rate of change.

After having presented above some interesting features of the GCSER we will investigate the possibility that the LSEFM is applicable on a unified global seismogenic area. Thus, we will examine "a posteriori" and in detail the GCSER for shorter periods of time before the occurrence of the mega - earthquakes shown in TABLE – 1. It is expected that in case the LSEFM is valid then the normal GCSER for each earthquake case will present the characteristic features (normal seismic energy flow, locked period of the seismogenic area) which enable the corresponding magnitude calculation. Following are presented in graph form the magnitude calculations for each case of the earthquakes of TABLE - 1. The inclined line indicates the normal GCSER as a function of time while the distance between the two horizontal lines indicate the energy charge of the seismogenic area. The red arrow indicates the time of occurrence of the related mega - earthquake. The magnitude determinations are presented in the following figures (13, 14, 15, 16, 17).

**EQ: M = 9.0, 1952**

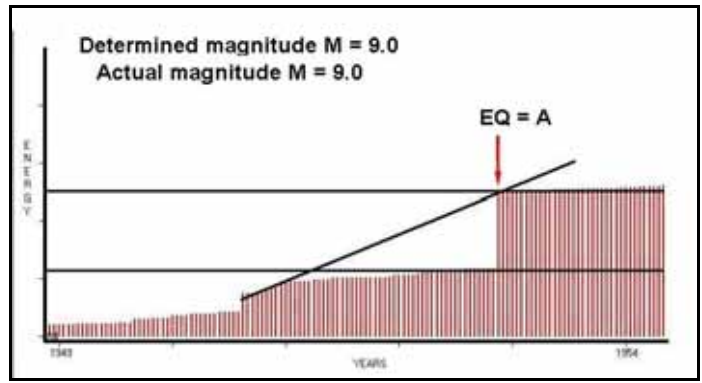

**Fig. 13. Determined magnitude M = 9.0 of 1952 EQ, compared to the actual (seismological) one of M = 9.0**

The normal GCSER of earthquake A is determined by the cumulative seismic energy release observed during the second half of year 1950. The intersection of the inclined line with a vertical line at the time of occurrence of the EQ defines the energy stored in the seismogenic area after subtracting the current seismic energy release just before the EQ occurrence. The determined stored energy is converted into EQ magnitude. In this case the determined magnitude and the actual (seismological) one are compatible.

**EQ: M = 9.5, 1960**

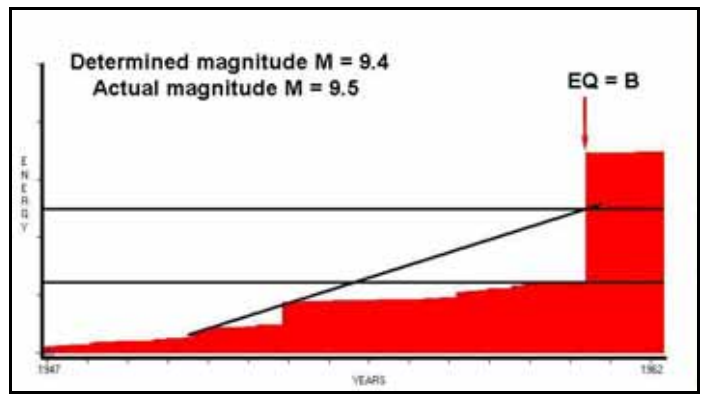

**Fig. 14. Determined magnitude M = 9.4 of 1960 EQ, compared to the actual (seismological) one of M = 9.5**

**EQ: M = 9.2, 1964**

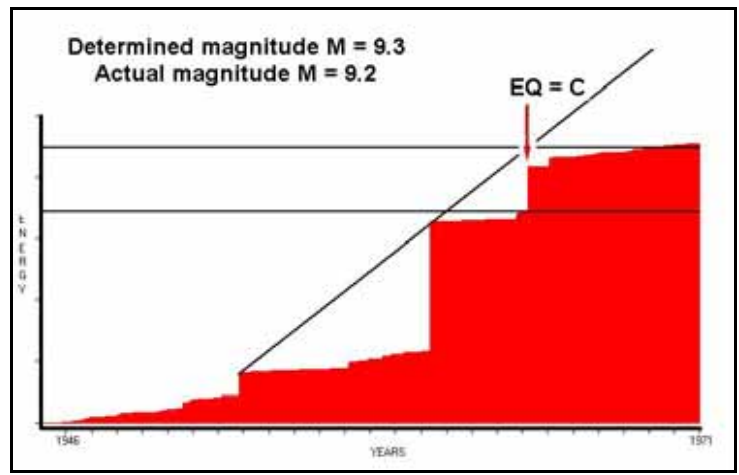

**Fig. 15. Determined magnitude M = 9.3 of 1964 EQ, compared to the actual (seismological) one of M = 9.2**

**In the above case the normal values for the GCSER were determined by the two (1952, 1960) past large seismic events.**

**EQ: M = 9.0, 2004**

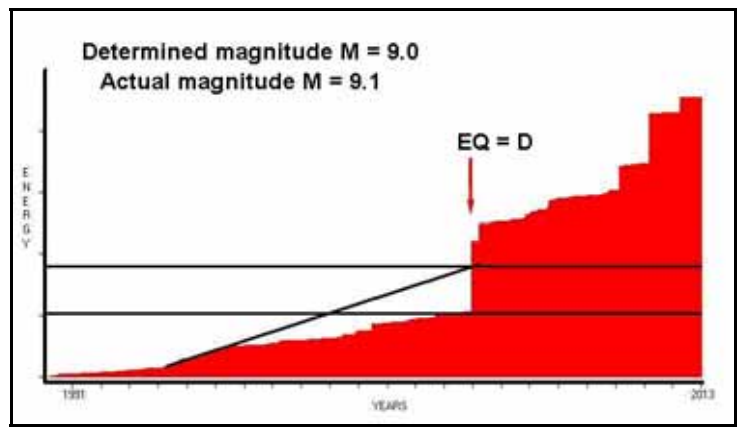

**Fig. 16. Determined magnitude M = 9.0 of 2004 EQ, compared to the actual (seismological) one of M = 9.1**

**EQ: M = 9.0, 2011**

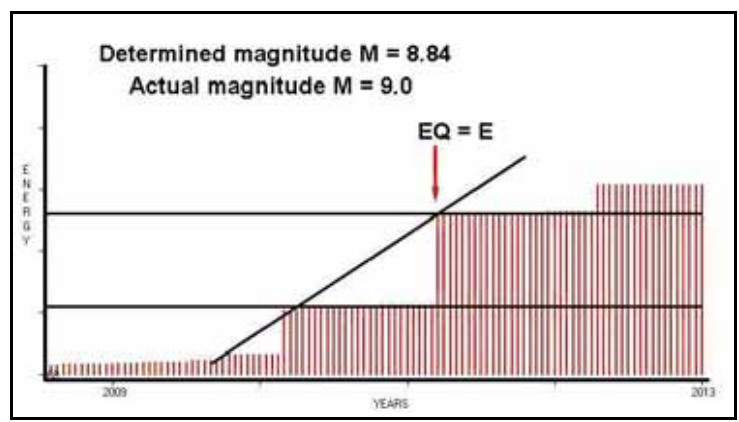

**Fig. 17. Determined magnitude M = 8.84 of 2011 EQ, compared to the actual (seismological) one of M = 9.0**

## 5. Discussion - Conclusions.

Studies on global seismicity have shown that observations indicate that, for great earthquakes, Earth behaves as a coherent seismotectonic system. A very large-scale mechanism for global earthquake triggering and/or stress transfer is implied. There are several candidates, but so far only viscoelastic relaxation has been modelled on a global scale (Bufe and Perkins, 2005). In another study it is demonstrated that the lithosphere does behave, at least in intermediate-term scale, as non-linear dynamic system that reveals classical symptoms of instability at the approach of catastrophe (Romashkova and Kossobokov, 2007). Furthermore, for a variety of magnitude cut-offs and three statistical tests, the global catalog, with local clusters removed, is not distinguishable from a homogeneous Poisson process. Moreover, no plausible physical mechanism predicts real changes in the underlying global rate of large events. Together these facts suggest that the global risk of large earthquakes is no higher today than it has been in the past (Shearer and Stark, 2011).

In contrast to the above mentioned works that are more or less based on statistical analyses, in the present work the global seismicity is treated entirely in terms of global cumulative seismic energy release. Therefore, all earthquakes are taken into account, in determining the energy release, no matter how they are characterized (foreshocks, main-shocks, aftershocks). The global cumulative seismic energy release (GCSER) is analyzed in the earthquake *magnitude domain* instead of being treated as an energy time series analyzed in the typical *frequency domain*. The main idea behind that analysis is that the GCSER is controlled by the magnitudes of the occurred EQs. Therefore, the magnitude spectrum of the global seismicity can be studied in details and specifically in specific energy (magnitude) windows. As an example we compare figure (4) that represents the GCSER for magnitudes from M = 2.0 to 10.0 for the period of time from 1900 to 2012, with figure (5) that represents the GCSER for M = 4.0 – 4.5 for the same period of time. It is clearly shown that the magnitude window for M = 4.0 – 4.5 is at a state of accelerating deformation for the last 50 years. For a specific magnitude window (fig. 6, M = 7.0 – 7.5) the GCSER is represented by a straight line. That is a definite indication that the global background seismicity is characterized by a magnitude of M = 7.0 – 7.5. The strain energy acquired by the lithosphere, globally, is balanced out by the background released seismicity of the latter magnitude. For larger earthquake magnitudes, the global seismicity is characterized by long periods of seismic quiescence in between the large successive earthquakes (figure 9).

A very interesting feature of GCSER, that is completely masked by seismic energy noise in figure (4), is its oscillating character. Its oscillation has been isolated in figure (12) while the five mega - earthquakes that occurred in the last 100 years are shown too. It is remarkable the timing of their occurrence in respect to the GCSER oscillation. The important question is: does any physical mechanism exist that accounts for the GCSER oscillation? Actually what is needed is a global oscillating lithospheric deformation of the same observed period of 50 – 60 years which may generate the required stress – strain oscillation, that in turn will activate the oscillating character of the global seismicity. Such mechanisms which are able to generate deformations of that extent have been proposed in the past.

Markowitz (1960) proposed a specific polar motion later called the "Markowitz wobble". The so-called Markowitz wobble (MW) is a quasi-harmonic variation of the mean pole of the Earth with a period of about 30 years and amplitude of 0.02"– 0.03" The origin of this motion still remains unknown. Dumberry and Bloxham (2002) and Dumberry (2008) developed a model to calculate the amplitude of the polar motion that results from an equatorial torque at the inner core boundary which tilts the inner core out of alignment with the mantle. They specifically addressed the issue of the role of the inner core tilt in the decade polar motion known as the Markowitz wobble. Gorshkov (2007) proposed a different mechanism. The North Atlantic Oscillation (NAO), which is characterized by large-scale phenomena in the system of atmosphere–ocean processes in this region, shows variations of some meteorological parameters in a wide frequency range. Synchronous oscillations of the pole (MW) and the NAO indices are revealed. The possibility of geophysical excitation of MW oscillations by variations of pressure fields in the North Atlantic was proposed.

Although physical mechanisms exist for a global scale oscillating lithospheric deformation, no one fits the period of about 60 years observed by the analysis of the GCSER. That misfit could be explained as follows: The global deformation is generated in the non linear system of the earth. Therefore, an activated fundamental oscillation could generate, due to global non-linear deformation, higher harmonics. In the reverse case the MW oscillation could be a higher harmonic oscillation with a period of 30 years whose its fundamental oscillation has a period of 60 years. The latter can very well be the topic of a future elaborated study.

The observed oscillation of the GCSER shows increasing amplitude in lime. Additionally the mega - earthquakes of the last 100 years did occur when the amplitude of the oscillation was increasing. The 5 mega - earthquakes of TABLE – 1 are clustered in two groups. The first one extends from 1952 to 1964 while the second, after a time interval of 40 years extends from 2004 to 2011. The clustering of the mega - earthquakes along the GCSER oscillation is characteristic.

What still remains in question is the mechanism that activates the amplitude increase of the GCSER oscillation. The MW driving mechanism is active since the formation of the earth in its present state. Therefore, the MW was always present despite the fact that it had not been studied earlier. Consequently, if only the MW is the triggering mechanism of mega - earthquakes, then similar EQs should have been observed in the past. The latter is not the case for the EQ data catalog which has been used for that study. Therefore, a mechanism must exist that will be capable to add an increasing stress – strain component to the oscillating one that is activated due to the MW driving mechanism.

In communication engineering systems two different signals are mixed by applying them in a nonlinear impedance system. The result of that operation is called *modulation* and the result is the change of the amplitude of the signal referred as "*carrier*" that is the signal that will carry the information of the second one. A typical example of modulation of a radio frequency "*carrier*" by an "*audio*" signal is presented (Orr, 1967) in figure (18).

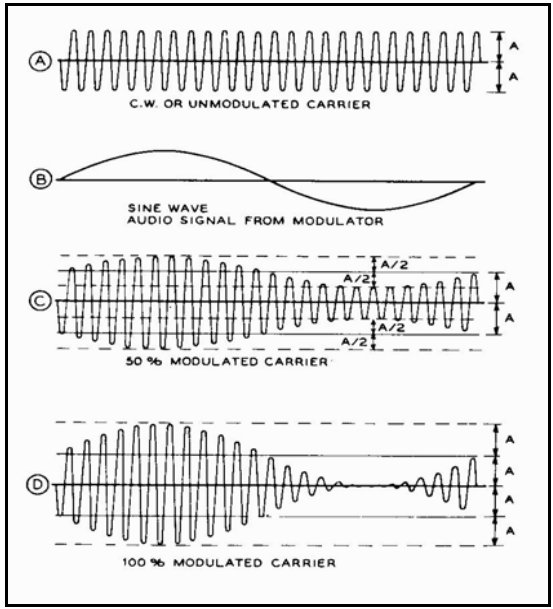

Fig. 18. Top drawing A represents an unmodulated carrier wave; B shows the audio modulating input signal; C shows the impressed on the carrier wave to the extent at 50% and D shows the carrier with 100% amplitude modulation (after Orr, 1967).

A similar very large scale mechanism could be activated for the case of global seismicity. The "*carrier*" is represented by the MW stress – strain lithospheric oscillation. What is additionally needed is the "*modulating*" signal. The latter could possibly be the stress – strain generated in the lithosphere due to the earth expansion as a result of the "Whole-Earth Decompression Dynamics" (Herndon, 2005). The nonlinear system needed for the amplitude modulation to activate is the earth itself. In the following figure (19) is shown the way the MW, the expanding earth and the earth's nonlinear system interact along in order to produce the resulting output.

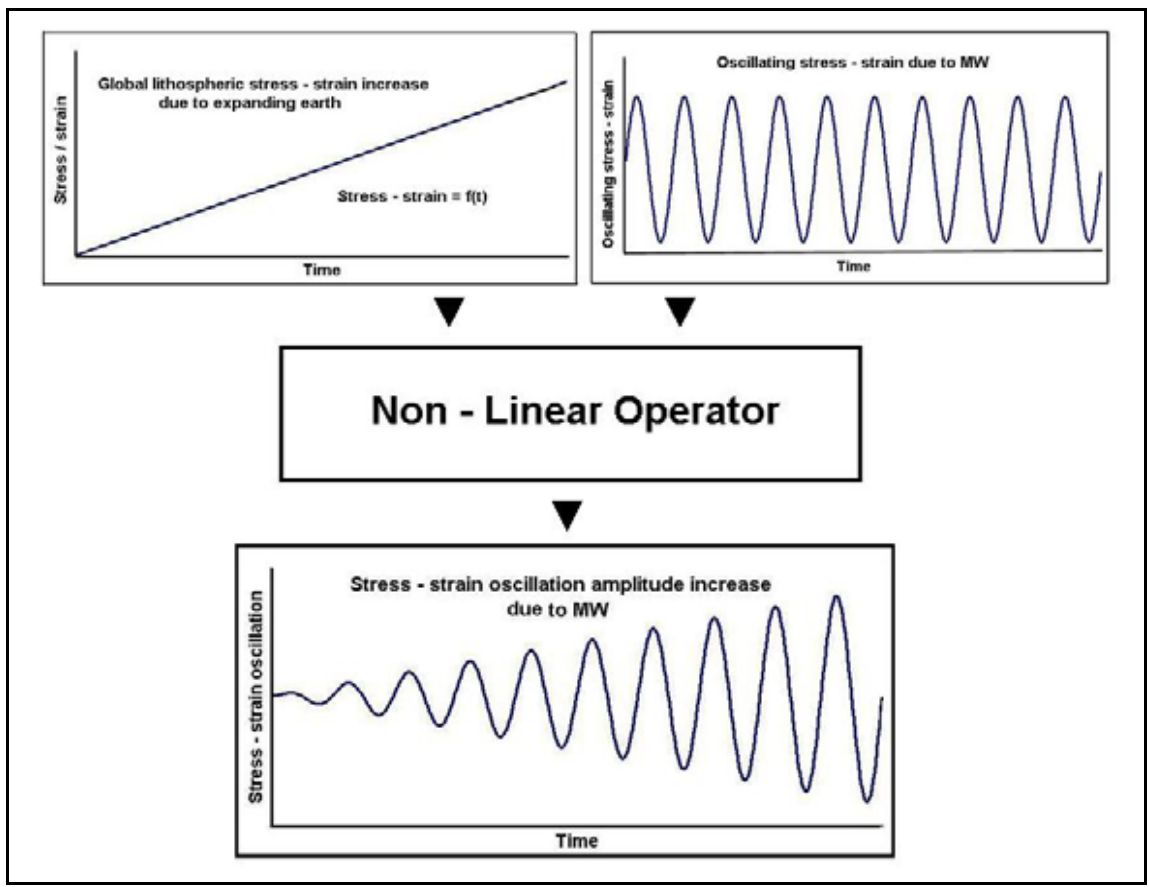

Fig. 19. **Upper left**: linear stress – strain increase induced in the lithosphere due to expansion of the earth. **Upper right**: Oscillating stress – strain lithospheric component with a period of 60 years due to MW hypothesized corresponding oscillation. **Middle**: Nonlinear earth system. **Lower**: amplitude modulated, by the earth expansion, MW stress – strain oscillation of the lithosphere.

The comparison of figure (19, bottom) to figure (12) suggests that the generating mechanism of the GCSER oscillation could very well be the one presented in figure (19) or some other very similar. It is evident that the mega - earthquakes that occurred during the present century are directly related to the amplitude increase of the GCSER oscillation.

The application of the Lithospheric Seismic Energy Flow Model (LSEFM) on the mega - earthquakes of the TABLE – 1 shows that the LSEFM is applicable on the global seismicity too. From figures (13) to (17) is shown that the global seismicity shows periods when it is at a state of balance, while for large periods it remains at a lock state. The magnitude of all five mega-earthquakes of TABLE - 1 has been calculated by applying the LSEFM methodology. In detail:

**EQ: M = 9.0, 1952:** the normal cumulative seismic energy flow (NCSEF) was determined by the first part of the inclined straight line. The calculated magnitude, **M = 9.0**, fits exactly the seismological one **M = 9.0**.

**EQ: M = 9.5, 1960:** the normal cumulative seismic energy flow (NCSEF) was determined by the first part of the inclined straight line and the next large seismic event (EQ: M = 9.0, 1952). The calculated magnitude, **M = 9.4**, fits quite well the seismological one **M = 9.5**.

**EQ: M = 9.2, 1964:** the normal cumulative seismic energy flow (NCSEF) was determined by the previous steps in GCSER denoted by the 1952 and 1960 seismic events. The calculated magnitude, **M = 9.3**, fits quite well the seismological one **M = 9.2**.

**EQ: M = 9.0, 2004:** the normal cumulative seismic energy flow (NCSEF) was determined by the first part of the inclined straight line. The calculated magnitude, **M = 9.0**, fits quite well the seismological one **M = 9.1**.

**EQ: M = 9.0, 2011:** the normal cumulative seismic energy flow (NCSEF) was determined by the first part of the inclined straight line and the next large seismic event (EQ: M = 8.8, 2010). The calculated magnitude, **M = 8.84**, is quite close to the seismological one **M = 9.0**.

The entire process indicates that large scale seismic quiescence seismic precursors mainly precede mega - earthquakes.

In conclusion, all the results can be summarized as follows:

- Global seismicity can be divided into the following classes:

   a. **Seismicity with M ≤ 7.0.** That type of seismicity indicates clearly an accelerating deformation character (fig. 5). The latter may indicate that a large seismic event is in preparation and may occur in the future.

   b. **Seismicity with M = 7.0 – 7.5** indicates that: the global background seismicity is at the level of M = 7.0 – 7.5, while the GCSER oscillates with a period of 60 years due to a large scale driving mechanism (fig. 19). Mega - earthquakes are controlled by the latter mechanism (fig. 10, 11, 12). The mega-earthquakes are closely related to the amplitude increase of the GCSER oscillation.

   c. **Seismicity with M = 7.5 – 8.0** shows a global "lock state" since 1923. The magnitude of the Mega-earthquakes A, B, C, D, and E of TABLE - 1 has been respectively calculated by the following figure (20) as M = 9.21, 9.29, 9.32, 9.51, and 9.53. Consequently the global lithospheric system is continuously strain-charged so that it is capable at any time to generate the next mega - earthquake. Moreover it is shown that its strain charge level continuously increases.

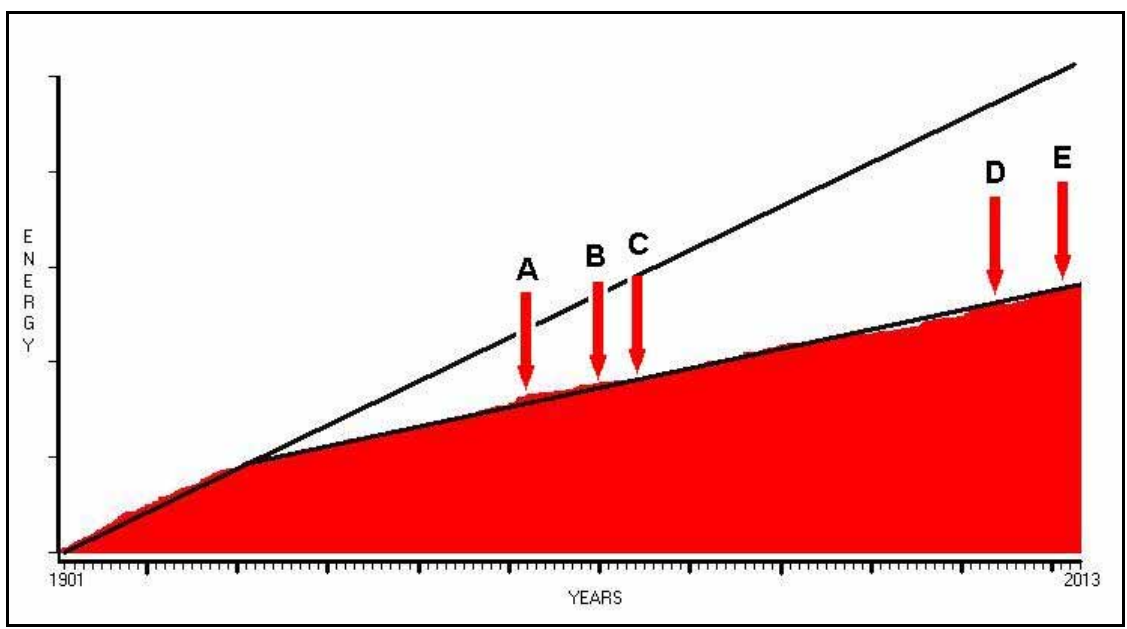

Fig. 20. The lithosphere, in a global scale, remains at a "lock state" since 1923. The red arrows indicate from left to right the 1952, 1960, 1964, 2004 and 2011 mega - earthquakes. The black lines indicate the normal cumulative seismic energy release for the corresponding periods (1900 – 1923 and 1923 – 2012). Used magnitude window M = 7.5 – 8.0

- Detailed information about the global seismicity evolution can be obtained by the use of the "magnitude / energy windows". Masked details of the global seismicity evolution in time can be extracted from the usual GCSER graph.

- The global seismicity, for the magnitude window of M = 7.0 – 7.5 shows an oscillatory character that is related to the generation of the mega - earthquakes of the recent century.

- The LSEFM model, tested on the 5 recent mega-earthquakes, found as been applicable on the global scale large seismicity.

- Taking into account the results of figure (12) and figure (20) a question could be raised concerning the future of the global seismicity. Shearer and Stark (2012) in a paper under the title: "Global risk of big earthquakes has not recently increased" by using statistical methods, addressed that question as: *"For a variety of magnitude cutoffs and three statistical tests, the global catalog, with local clusters removed, is not distinguishable from a homogeneous Poisson process. Moreover, no plausible physical mechanism predicts real changes in the underlying global rate of large events. Together these facts suggest that the global risk of large earthquakes is no higher today than it has been in the past".*
The latter statement is in complete disagreement to the findings of the work already presented. A global physical mechanism (fig. 19) has been postulated that accounts for the continuous increase of the global lithospheric stress – strain charge. The two recent (2004, 2011) mega-earthquakes justify the postulated global mechanism.

- As a result, mega - earthquakes are <u>more probable to occur in the future</u> due to the postulated mega - earthquakes activating mechanism, as shows the amplitude increase of the oscillating GCSER.

## 6. References.